\documentclass[11pt,twoside]{article}
\usepackage{asp2014}

\aspSuppressVolSlug
\resetcounters

\begin{document}

\title{The use of Specutils by Data Central}

\author{James~Tocknell$^1$}
\affil{$^1$Australian Astronomical Optics - Macquarie, Faculty of Science and
Engineering, Macquarie University, North Ryde, NSW 2113, Australia; \email{james.tocknell@mq.edu.au}}

\paperauthor{James~Tocknell}{james.tocknell@mq.edu.au}{0000-0001-6637-6922}{}{Australian
  Astronomical Optics - Macquarie, Faculty of Science and Engineering, Macquarie
  University}{North Ryde}{NSW}{2113}{Australia}



  
\begin{abstract}

Specutils is an Astropy affiliated package which provides a consistent interface
to astronomical spectra (primarily 1D). As Specutils can be adapted to parse
spectra in many different formats, Specutils plays a key role at Data Central,
allowing us to handle the diverse formats provided to us by survey teams. In
this poster, I will cover what Specutils is, how it works, how Data Central uses
it, and why you too should use and contribute to it.
  
\end{abstract}

\section{What is Specutils}

Specutils \citep{2019ascl.soft02012A} is an Astropy coordinated python package,
that is, it is centrally maintained by the Astropy Project
\citep{2013A&A...558A..33A,2018AJ....156..123A,2022ApJ...935..167A}.
Specutils provides three key classes:
\texttt{Spectrum1D}, \texttt{SpectrumCollection}, and
\texttt{SpectrumList}\footnote{Documentation for these classes, as well as the
  rest of Specutils, can be found at \url{https://specutils.readthedocs.io/}};
these are built on core Astropy classes, allowing the usual interaction with
arithmetic, slicing and units.

Specutils provides numerous routines that consume these classes, providing such
features as: line and continuum fitting; smoothing, convolution and resampling;
template matching and uncertainty estimation; and reddening of model spectra.

Specutils also ties into the Astropy IO registry system\footnote{
  Details about the Astropy IO registry system can be found at
  \url{https://docs.astropy.org/en/stable/io/registry.html}.
}, providing a user-friendly interface to read and write the various formats
spectra are stored in. There is support for more generic formats such as ASCII
tables or CSV, but also for more specific formats such as IRAF's FITS, and
instrument or survey-specific formats such as HST, Subaru, SDSS and MaNGA. By
using the Astropy IO registry system, significant proportions of a codebase can
be written to be independent of the underlying file format, reducing its
maintenance cost.

\section{How Data Central uses Specutils}
Data Central \citep{2020SPIE11203E..1DO,2019lgei.confE..11M} has added and continues to add support for reading spectra from
both the hosted surveys and the telescope archives stored at Data Central. We do
this for two reasons:
\begin{enumerate}
  \item This reduces the barrier to entry to using this existing
    data, especially data from older surveys where the data format may have
    incorrect FITS WCS or where the format varies across the whole survey.
  \item Data Central can build common tools to analyse and visualise this
    spectra, rather than needing to build bespoke tools, modify the original
    data, or limit ourselves to specific formats. Given Data Central hosts a
    number of legacy surveys from the Anglo-Australian Telescope, this ability
    to adapt to different formats but provide the same interface is vital for
    us. See fig.~\ref{P27_f1} for an example tool we developed using loaders we
    have contributed to Specutils.
\end{enumerate}

\articlefigure{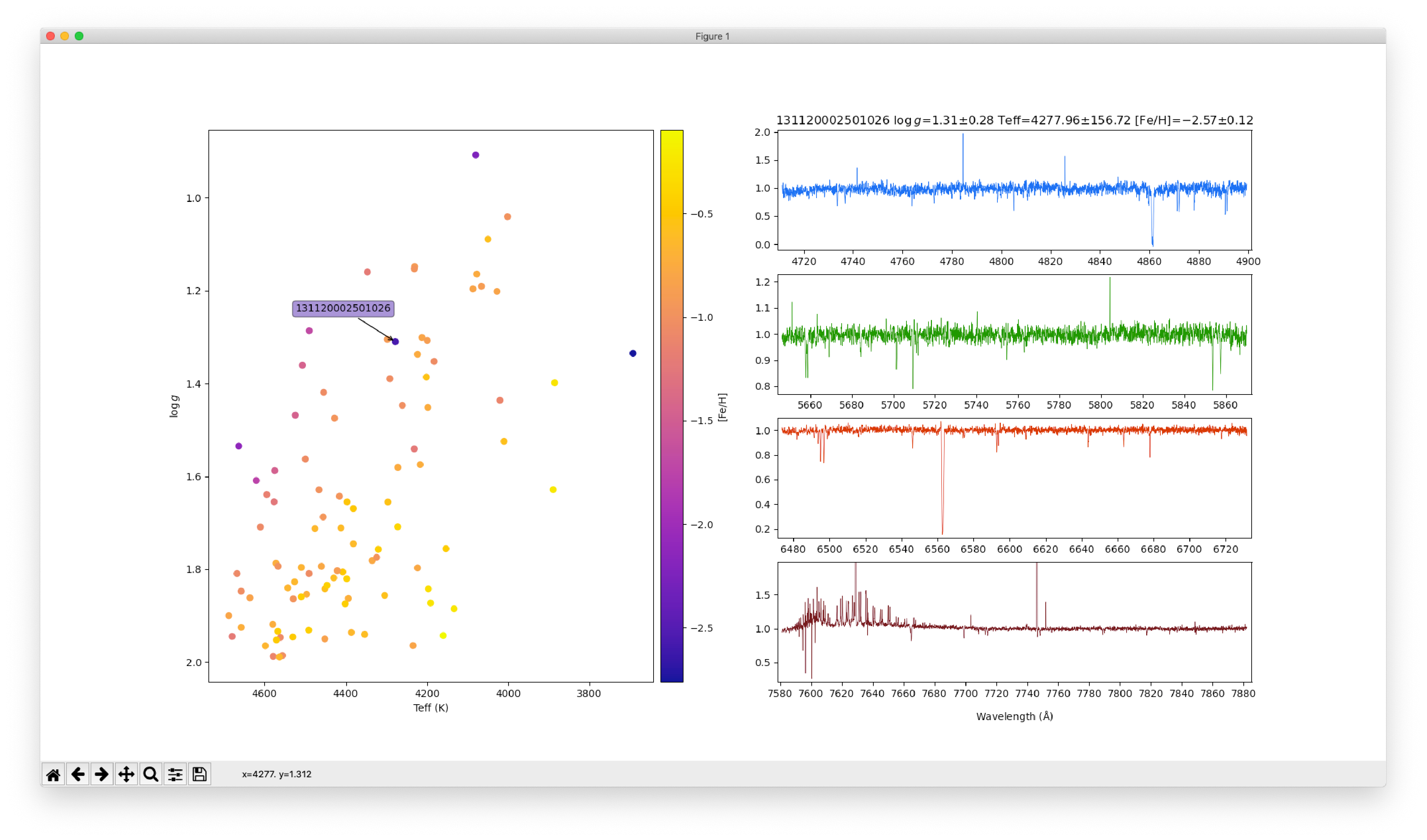}{P27_f1}{A screenshot of an example application for
  browsing GALAH spectra.
  Further examples can be found in the Data Central documentation at
  \url{https://docs.datacentral.org.au/}.
}

\section{Contributing to Specutils}

\citet{C23_adassxxxii} has a detailed discussion of why archives should look at
value-add services, and writing Specutils loaders for your spectral data should
be one of those.  For those whose systems are written in Python, this will
reduce the time spent on developing new tools, and allow adaption of others
which build upon Specutils. Even if your systems are not written in Python, many
of your users will be using Python \citep{2015arXiv150703989M},
so writing Specutils loaders will increase the value of your archive.

Contributing to Specutils is done in the same way as other Astropy projects:
there is a GitHub repository to which you can send Pull Requests. The Astropy
provides detailed contribution instructions in case you are not yet familiar
with the contribution process.

\section{Summary}

We strongly encourage those who host spectra, especially those who existing
infrastructure is built on top of Python, to contribute support for your
spectra: this will reduce the barrier to Astronomers to using your data, and the
contribution process to Specutils is not onerous.

\bibliography{P27}


\end{document}